\documentclass[11pt]{article}

\usepackage{amsmath,amssymb,amsthm}
\usepackage[numbers]{natbib}
\usepackage{graphicx}
\usepackage{booktabs}
\usepackage{geometry}
\usepackage{bm}
\usepackage{enumitem}
\usepackage{caption}
\usepackage{subcaption}
\usepackage{float}
\usepackage{xcolor}
\usepackage{tikz}
\usepackage{array}
\usepackage{xcolor}

\usepackage[numbers]{natbib}
\definecolor{deepblue}{RGB}{0,70,140}
\definecolor{tealgreen}{RGB}{0,120,110}
\definecolor{darkred}{RGB}{150,40,40}
\usepackage[
    colorlinks=true,
    linkcolor=darkred,
    citecolor=tealgreen,
    urlcolor=deepblue
]{hyperref}

\usetikzlibrary{arrows.meta,positioning,fit,shapes.geometric,calc}

\renewcommand{\toprule}{\hline\hline}
\renewcommand{\midrule}{\hline}
\renewcommand{\bottomrule}{\hline}

\title{%
Quasi-material finite-time rotationally coherent sets in photospheric supergranulation
}

\author{%
Francisco J. Beron-Vera\\
Department of Atmospheric Sciences\\
Rosenstiel School of Marine, Atmospheric, and Earth Science\\
University of Miami\\
Miami, Florida, USA\\
\href{mailto:fberon@miami.edu}{\texttt{fberon@miami.edu}}
}

\date{Started: June 7, 2026.  This version: \today.}

\begin{document}

\maketitle

\begin{abstract}
    Supergranular flows organize transport in the solar photosphere over spatial and temporal scales much larger than granulation. While coherent vortical motions have been identified using objective Lagrangian diagnostics such as the Lagrangian-averaged vorticity deviation (LAVD), rotational coherence captures only one aspect of coherent flow organization. Here we introduce finite-time rotationally coherent sets (FTRCS) by combining the inflated dynamic Laplacian (IDL), which identifies finite-time quasi-material coherent regions, with LAVD-based rotational diagnostics. The IDL extracts coherent structures with finite lifetimes, while LAVD identifies those exhibiting enhanced intrinsic rotation. Application to photospheric velocity fields shows that instantaneous vortical features do not necessarily correspond to finite-time rotationally coherent structures. The analysis also illustrates the effect of compressibility: coherent sets may form through persistent contraction associated with convergent transport, rather than through the persistence of rotating material regions. The combined IDL--LAVD approach separates finite-time transport coherence from intrinsic rotational organization in time-dependent flows.
\end{abstract}


\tableofcontents

\section{Introduction}

The solar photosphere contains a hierarchy of convective motions ranging from granulation to supergranulation. Supergranular flows form cellular patterns with characteristic horizontal scales of tens of megameters and lifetimes of order one day. These flows organize the transport of magnetic flux, contribute to the formation of the magnetic network, and influence the dynamics of the solar atmosphere. Numerical models of photospheric magnetoconvection reproduce persistent flow and magnetic structures arising from the interaction of convection and magnetic fields \cite{Vogler-etal-24}.

The Lagrangian analysis of photospheric flows has recently attracted attention because coherent structures identified from trajectory evolution reveal transport properties that are not apparent from instantaneous velocity fields. Finite-time Lagrangian diagnostics have been used to identify transport barriers and organizing structures in supergranular flows, highlighting the role of material transport in the formation of photospheric patterns \cite{Chian-etal-19}. A related class of objective diagnostics based on the \emph{Lagrangian-averaged vorticity deviation} (LAVD) \cite{Haller-etal-16} identifies regions undergoing coherent material rotation. These methods have revealed rotationally coherent structures in solar and astrophysical flows \cite{Rempel-etal-17}. Silva et al.~\cite{Silva-etal-20} applied LAVD methods to supergranular flows, showing that coherent vortices can be extracted from photospheric velocity data.

Supergranulation, however, is not primarily a vortex-dominated process. The dominant cellular motions involve strong divergence and convergence, and regions that organize transport need not coincide with rotationally coherent vortices. This motivates separating finite-time transport coherence from rotational coherence.

The dynamic Laplacian provides a variational approach for identifying finite-time coherent sets by seeking regions whose boundaries remain small under material evolution \cite{Froyland-15, Froyland-26}. Classical formulations identify material sets that remain coherent over a prescribed analysis interval, effectively selecting structures that persist throughout that time span. This requirement can be restrictive in flows where coherent structures appear, merge, split, or disappear during the observation window, as occurs in many geophysical and astrophysical systems.

The \emph{inflated dynamic Laplacian} (IDL) \cite{Froyland-Koltai-23, Froyland-26} extends this approach by embedding time as an additional coordinate and introducing an inflation-dependent penalty for departures from material evolution. This allows the identification of quasi-material coherent sets with finite lifetimes, rather than requiring persistence over the entire observation window. Related approaches have been used to study the formation and destruction of oceanic and atmospheric coherent structures \cite{Andrade-etal-20, Andrade-etal-25, Atnip-etal-24}. In nearly incompressible flows, finite lifetimes are typically associated with splitting, merging, formation, or disappearance of coherent regions. In strongly compressible flows such as photospheric supergranulation, coherent sets may also result from contraction toward convergent regions, leading to a different interpretation of the extracted structures.

In this work, we combine IDL and LAVD to define \emph{finite-time rotationally coherent sets} (FTRCS). The IDL is first used to identify finite-time quasi-material coherent regions, including those produced by compressible transport. The LAVD is then evaluated within these regions to determine which ones exhibit enhanced intrinsic rotation. This approach provides an IDL-based construction of rotationally coherent sets in which transport coherence and rotational coherence are treated as separate properties.

The remainder of the paper is organized as follows. Section~\ref{lab:ftrcs} introduces FTRCS by combining finite-time quasi-material coherence from the IDL with intrinsic material rotation measured by the LAVD. Section~\ref{lab:applications} applies the method to photospheric supergranular velocity fields. Section~\ref{lab:conclusions} summarizes the main findings. The mathematical background of material coherent sets and the dynamic Laplacian, together with details of the inflated formulation and numerical implementation, are provided in Appendices~\ref{app:dl} and~\ref{app:idl}.

\section{Finite-time rotationally coherent sets}
\label{lab:ftrcs}

Identifying rotationally coherent structures over finite time intervals requires separating two properties: transport coherence and rotational coherence. Transport-coherent regions remain organized under material evolution, with limited boundary filamentation, whereas rotationally coherent regions exhibit enhanced intrinsic material rotation. A coherent material region may translate, deform, expand, or contract with little rotation, and regions of large instantaneous vorticity need not form coherent material structures.

We define FTRCS by combining these two diagnostics. Finite-time transport coherence is obtained from the IDL of Froyland and Koltai \cite{Froyland-Koltai-23}, while rotational content is evaluated independently using the LAVD of Haller et al.~\cite{Haller-etal-16}.

Let \(M\subset\mathbb R^2\) denote the spatial domain and let \(F_{t_0}^{t}:M\to M_t\) be the finite-time flow map generated by the velocity field, mapping an initial position \(x\in M\) to its position at time \(t\).

The IDL extends the dynamic Laplacian formulation of material coherent sets introduced by Froyland \cite{Froyland-15} to structures with finite lifetimes. Instead of requiring a coherent material region to persist throughout the complete analysis interval under the flow map \(F_{t_0}^{t}\), the IDL searches for coherent structures in an inflated space-time domain. These structures are \emph{quasi-material}: their support may evolve within the observation window, while departures from material evolution are penalized.

The IDL eigenfunctions provide a spectral relaxation of this space-time coherence problem. As detailed in Appendices~\ref{app:dl} and~\ref{app:idl}, the inflated formulation leads to the eigenvalue problem associated with
\begin{equation}
   \Delta_a^D
   =
   \Delta^D
   +
   a^2\partial_{tt},
\end{equation}
where \(\Delta^D\) is the dynamic Laplacian and the additional term controls departures from material evolution in the inflated time direction. Coherent sets are extracted from the eigenfunctions associated with the smallest nonzero eigenvalues, whose nearly constant regions provide relaxed approximations of coherent-set indicators.

The corresponding discrete problem represents a function on the inflated space-time domain by a vector \(w\). The Rayleigh quotient associated with the discretized operator can be written as
\begin{equation}
   \rho_a(w)
   =
   \frac{w^\top K_{\rm space}w}{w^\top Gw}
   +
   a^2
   \frac{w^\top K_{\rm material}w}{w^\top Gw}.
   \label{eq:rhoa}
\end{equation}
Here \(K_{\rm space}\) and \(K_{\rm material}\) are the discrete contributions associated with spatial gradients and departures from material evolution, respectively, and \(G\) is the mass (Gram) matrix defining the discrete inner product. The inflation parameter \(a\) controls the relative weight of these two effects. Larger values of \(a\) penalize departures from material evolution more strongly, approaching the dynamic Laplacian description of persistent material coherent sets.

The IDL computation does not include rotational information. To determine which IDL coherent sets also display organized rotation, we evaluate the LAVD along trajectories. For \(x_t=F_{t_0}^{t}(x)\), the LAVD over \([t_0,t_1]\) is
\begin{equation}
   \mathrm{LAVD}_{t_0}^{t_1}(x)
   :=
   \int_{t_0}^{t_1}
   \left|
   \omega(F_{t_0}^{t}(x),t)
   -
   \langle\omega\rangle(t)
   \right|
   dt ,
\end{equation}
where \(\omega(x,t)=\hat{\mathbf z}\cdot\nabla\times\mathbf u(x,t)\) is the vertical vorticity and \(\langle\omega\rangle(t)\) is its spatial mean. Subtracting the spatial mean removes the contribution from time-dependent rigid-body rotations of the reference frame, so that LAVD measures accumulated intrinsic material rotation.

The standard LAVD approach identifies rotationally coherent vortices from closed convex contours surrounding LAVD maxima. Here LAVD is used only as a diagnostic after the IDL calculation. It does not determine the coherent sets, but measures the rotational content of the quasi-material regions obtained from the IDL.

The procedure is:
\begin{enumerate}[label=(\roman*),nosep]
   \item Identify quasi-material coherent sets using the IDL.
   \item Compute the LAVD field over the same time interval.
   \item Evaluate the LAVD distribution inside each IDL coherent set relative to the background flow.
\end{enumerate}

An IDL coherent set is classified as a finite-time rotationally coherent set when its accumulated intrinsic rotation is enhanced relative to the surrounding domain. The classification keeps transport coherence and rotational coherence as separate quantities, making it possible to distinguish coherent regions that rotate from those dominated by translation, deformation, or contraction.

\section{Application to supergranular flow}\label{lab:applications}

We apply the FTRCS detection procedure to a two-dimensional photospheric velocity field representative of quiet-Sun supergranular dynamics. The data set, obtained from observations reported by Go\v{s}i\'{c} et al.~\cite{Gosic-etal-14} and previously analyzed from a Lagrangian coherent structure viewpoint by Chian et al.~\cite{Chian-etal-19}, consists of horizontal plasma velocities derived from continuum intensity images acquired near the solar disc center by the Narrowband Filter Imager onboard the \emph{Hinode} satellite as part of \emph{Hinode} Operation Plan 151 \cite{Kosugi-etal-07}. The velocity fields were obtained using local correlation tracking (LCT) of the intensity images after removal of five-minute oscillations, with the processing described in Requerey et al.~\cite{Requerey-etal-18}. Supergranulation is characterized by cellular convective motions with typical horizontal scales of \(20\)--\(30\,{\rm Mm}\) and lifetimes of order one day. The velocity field is represented on a fixed square observational domain \(M=[0,L]^2\), with \(L\approx5\times10^4\,{\rm km}\), resolved on a \(432\times432\) grid with spatial resolution \(116\,{\rm km}\) and temporal spacing \(0.025\,{\rm h}\). Material evolution is described by the finite-time flow maps \(F_{t_0}^t\), whose images need not coincide with \(M\) because the observed horizontal photospheric flow is compressible.

\begin{figure}[t!]
    \centering
    \includegraphics[width=\textwidth]{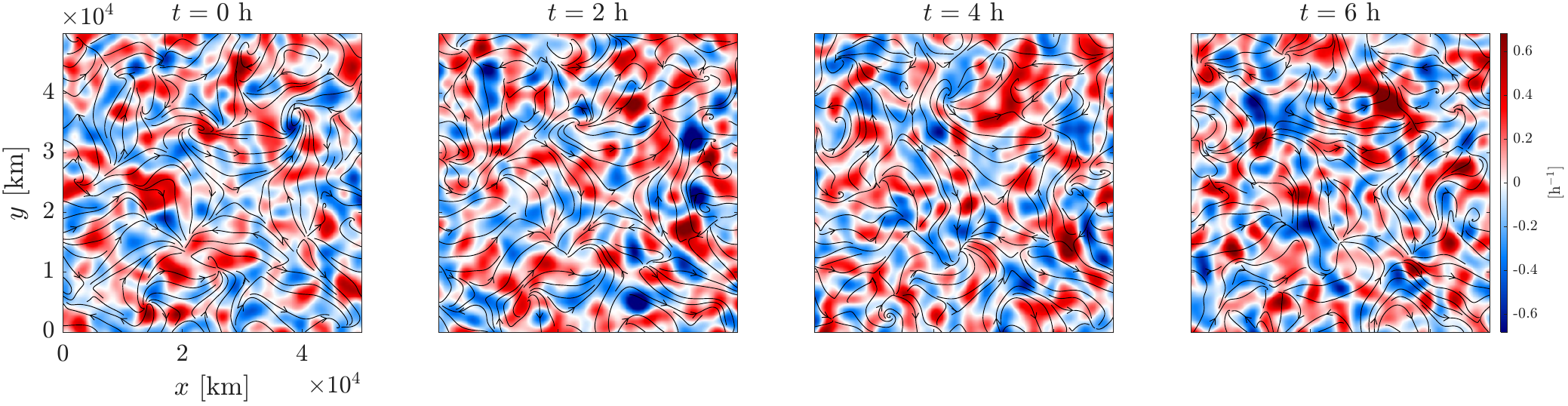}
    \caption{Snapshots of the supergranular velocity field. Colors show the instantaneous vorticity, and black curves indicate streamlines.}
    \label{fig:vort_stream}
\end{figure}

Figure~\ref{fig:vort_stream} shows snapshots of the velocity field using instantaneous streamlines overlaid on the vertical vorticity. These Eulerian diagnostics illustrate the characteristic cellular organization of supergranular dynamics, including divergent outflows from cell interiors, convergence near cell boundaries, strong deformation, and localized vorticity. However, instantaneous streamline patterns and regions of large vorticity may appear, disappear, or deform over time without defining persistent material objects. The IDL construction instead identifies coherent sets through the geometry of their finite-time material evolution under \(F_{t_0}^t\).

We compute IDL modes over the interval \(0\le t\le6\) h. Although the photospheric velocity field is compressible, the corresponding flow-map diagnostic shows that trajectory evolution over this interval is dominated by finite-time contraction into interior attracting regions rather than by loss through the computational boundary. This supports the use of this time window for coherent-set analysis.

The inflation parameter \(a\) controls the relative weighting between spatial compactness and temporal coherence. Small values of \(a\) favor sets with small instantaneous spatial boundary, whereas larger values increasingly penalize departures from material evolution. Since \(a\) represents a scale-selection parameter rather than a quantity determined directly from the data, we analyze a range of values rather than prescribing it a priori.

\begin{figure}[t!]
    \centering
    \begin{subfigure}{0.49\textwidth}
        \includegraphics[width=\textwidth]{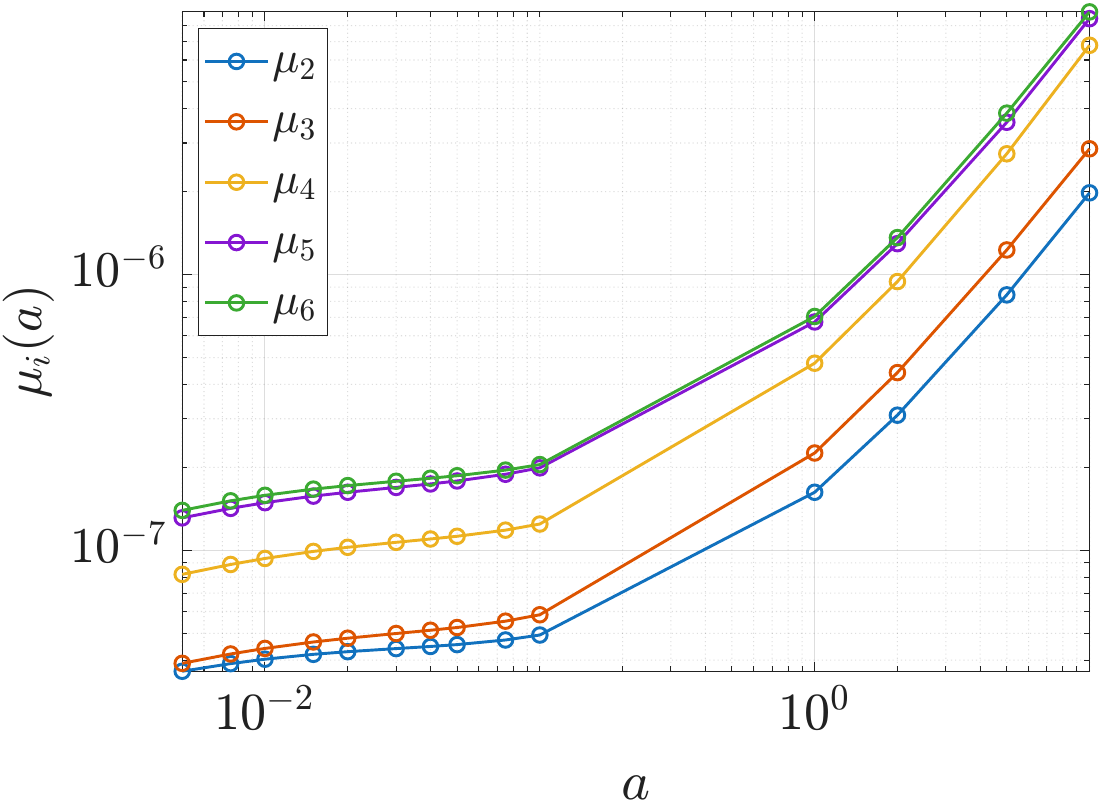}
        \caption{}
    \end{subfigure}
    \begin{subfigure}{0.49\textwidth}
        \includegraphics[width=.975\textwidth]{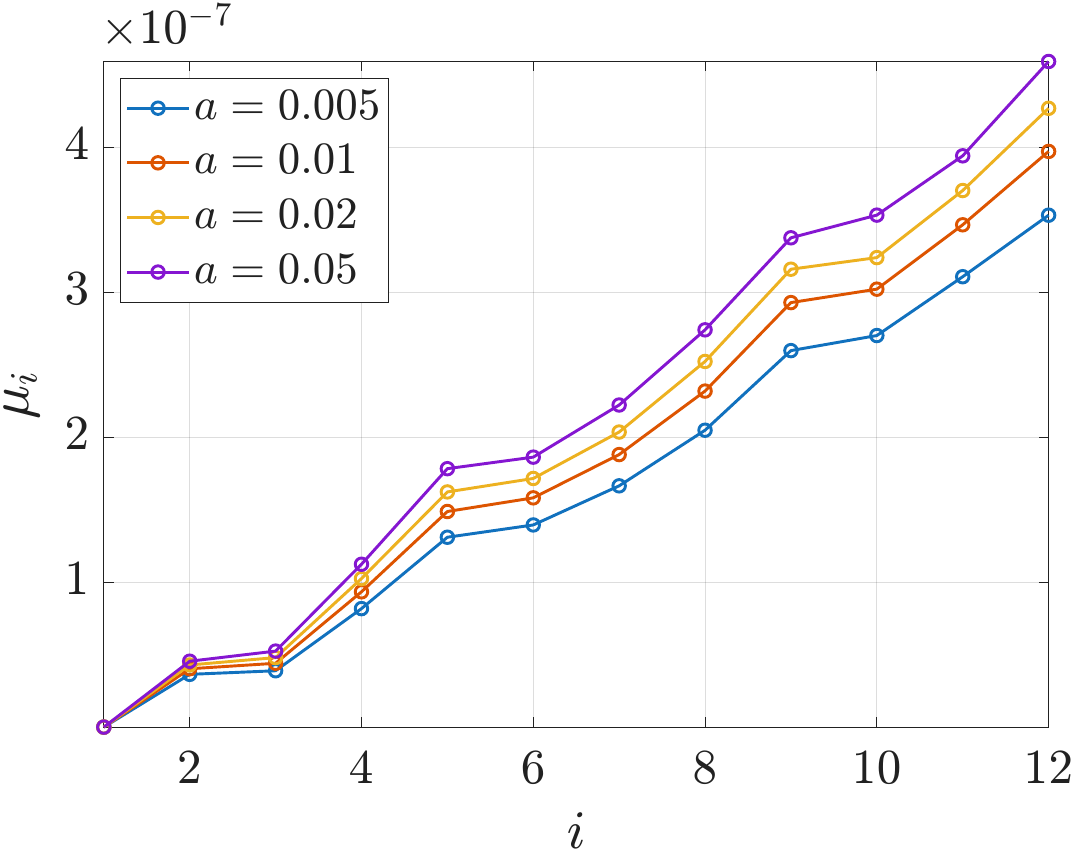}
        \caption{}
    \end{subfigure}
    \caption{IDL spectral diagnostics. (a) First nontrivial generalized eigenvalues \(\mu_i(a)=-\lambda_i(a)\) as functions of the inflation parameter \(a\), excluding the constant mode. (b) Eigenvalue distributions as functions of mode index for selected values of \(a\).}
    \label{fig:idl_spectrum}
\end{figure}

Figure~\ref{fig:idl_spectrum} summarizes the dependence of the IDL spectrum on the inflation parameter \(a\). Figure~\ref{fig:idl_spectrum}a shows the first nonzero eigenvalues \(\mu_i(a)\) of the positive IDL operator (equivalently, \(\mu_i=-\lambda_i\) for the Laplacian convention \(\Delta_a^D f_i=\lambda_i f_i\)), excluding the constant mode, as \(a\) is varied. For small values of \(a\), the leading eigenvalues vary weakly, consistent with a regime in which the spatial dynamic-Laplacian contribution dominates. As \(a\) increases, the material-derivative penalty becomes more influential and the eigenvalues change accordingly. The spectrum therefore provides a useful global diagnostic of the dependence on \(a\), but it does not by itself identify a unique transition value or an optimal inflation scale.

Figure~\ref{fig:idl_spectrum}b shows the corresponding eigenvalue distributions as a function of mode index for selected values of \(a\). The spectra do not exhibit a pronounced spectral gap separating a small number of dominant coherent structures from the remaining modes. The absence of a clear gap persists when additional modes are computed, so the spectrum does not provide a unique choice for the dimension of the coherent eigenspace. We therefore interpret the leading eigenvectors as defining a coherent subspace rather than a unique spectral partition of the flow. In the computations below, we retain the leading 12 IDL modes, discard the constant mode, and use the remaining 11-dimensional nonconstant eigenspace for coherent-set extraction.

Following Froyland et al.~\cite{Froyland-etal-19}, we apply Sparse EigenBasis Approximation (SEBA) to the leading IDL eigenspace. Let \(V=[v_1,\ldots,v_k]\) denote the matrix whose columns are the coefficient vectors of the first \(k\) nonconstant IDL eigenfunctions retained after the spectral calculation. SEBA constructs localized vectors \(S=[s_1,\ldots,s_k]\) by approximately rotating the eigenspace spanned by the columns of \(V\). This step is important when individual eigenfunctions do not correspond to individual coherent structures, as occurs when several modes have comparable coherence. The resulting vectors \(s_j\) behave as sparse approximate indicator functions of coherent-set candidates, but are generally not eigenvectors of the IDL operator.

SEBA produces sparse real-valued functions rather than binary characteristic functions. We therefore convert each \(s_j\) into a space-time support by retaining the region where \(s_j\) exceeds the 90th percentile of its positive values and keeping the largest connected component. Components with fewer than 20 grid cells are discarded as unresolved numerical fragments.

To diagnose the effect of the inflation parameter on the extracted candidates, we evaluate the Rayleigh quotient \eqref{eq:rhoa} for each SEBA vector \(s_j\). Since the SEBA vectors are localized combinations of IDL eigenvectors, this provides a candidate-by-candidate measure of the balance between spatial compactness and material persistence. We report the relative contributions
\begin{equation}
   r_{\rm space}
   :=
   \frac{\rho_{\rm space}(s_j)}
   {\rho_a(s_j)},
   \qquad
   r_{\rm material}
   :=
   \frac{a^2\rho_{\rm material}(s_j)}
   {\rho_a(s_j)} ,
\end{equation}
which satisfy \(r_{\rm space}+r_{\rm material}=1\).

These quantities are used only as diagnostics of the inflation regime: the IDL method itself does not require equality of the two contributions, and the crossover is not a spectral selection criterion. No filtering based on \(\rho_a(s_j)\) is applied when defining the coherent-set candidates.

\begin{figure}[t!]
   \centering
   \includegraphics[width=0.75\textwidth]{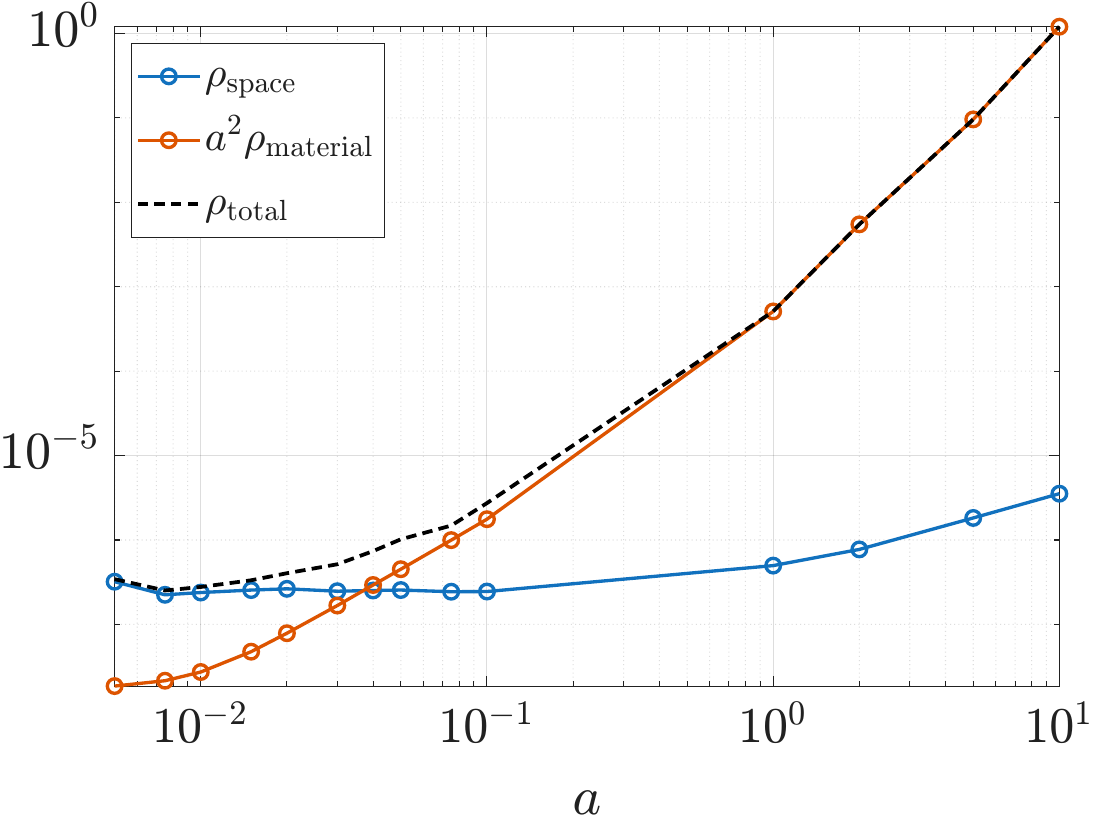}
   \caption{Decomposition of the IDL-SEBA Rayleigh quotient into spatial and material contributions. Curves show median values over the 11 SEBA candidates obtained from the nonconstant IDL eigenspace.}
   \label{fig:rayleigh}
\end{figure}

Figure~\ref{fig:rayleigh} provides a more direct diagnostic of the inflation regime by decomposing the SEBA-candidate Rayleigh quotients into spatial and material contributions. For each value of \(a\), the 11 SEBA candidates obtained from the nonconstant IDL eigenspace are evaluated using \(\rho_a(s_j)\). The plotted quantities show median values of \(\rho_{\rm space}(s_j)\), \(a^2\rho_{\rm material}(s_j)\), and \(\rho_a(s_j)\) over these localized candidates. These curves therefore characterize the typical balance between spatial compactness and material persistence of the extracted coherent-set candidates, rather than the IDL spectrum itself.

For small \(a\), the Rayleigh quotient is controlled primarily by spatial compactness. As \(a\) increases, the contribution associated with the material derivative grows and eventually dominates the inflated geometry. The crossover between these contributions provides a practical range of values where material persistence affects the extracted structures without overwhelming the spatial contribution. We select \(a=0.03\) as a representative value in this transition regime.

\begin{figure}[t!]
    \centering
    \includegraphics[width=\textwidth]{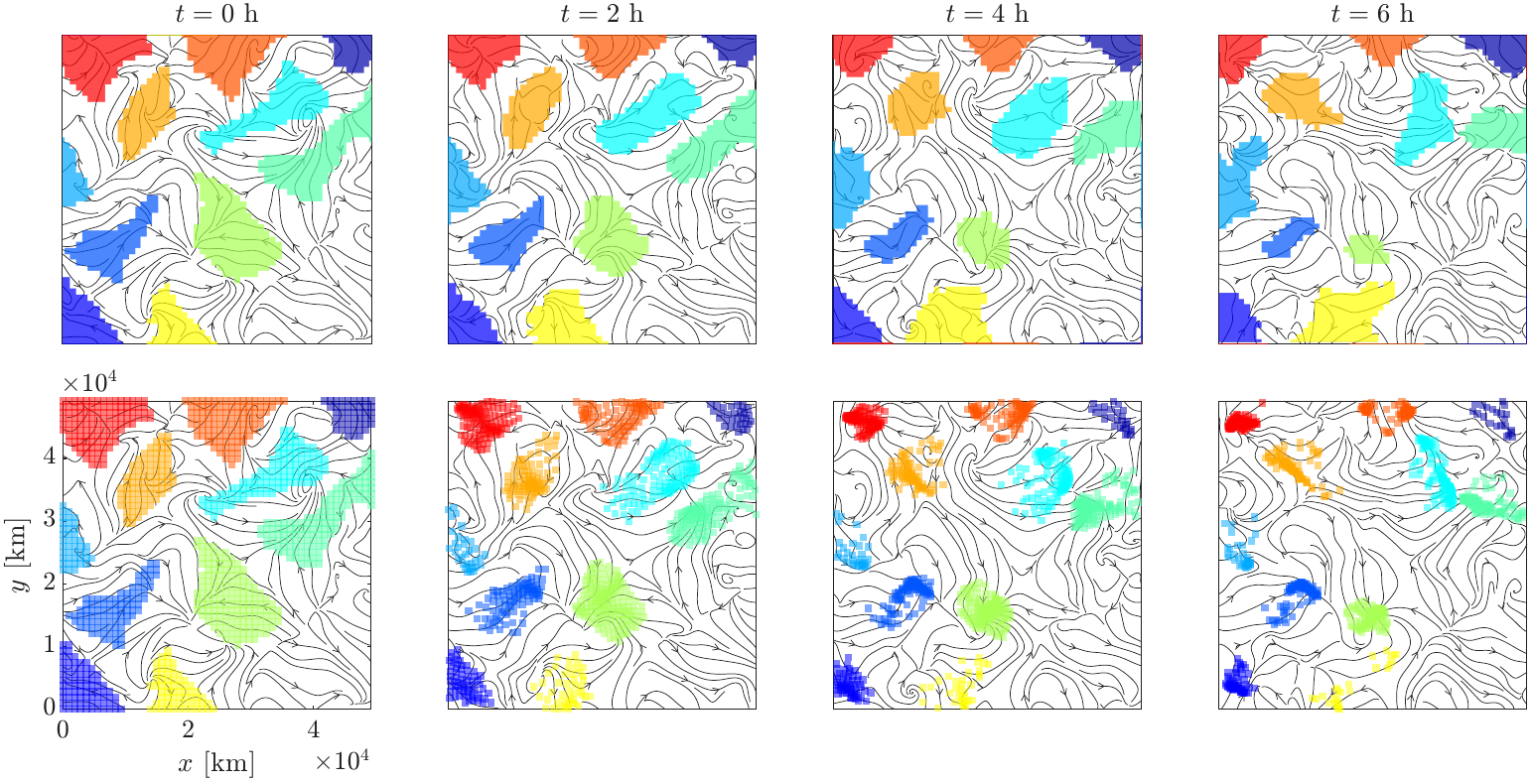}
    \caption{Materiality diagnostic for IDL-SEBA coherent-set candidates. The top row shows IDL supports at each time, and the bottom row shows the corresponding advected initial supports under the supergranular flow.}
    \label{fig:idl_materiality}
\end{figure}

Figure~\ref{fig:idl_materiality} compares the IDL-SEBA candidates with direct material advection by the observed velocity field. Let \(A_j(t)\subset M_t\) denote the support of the \(j\)-th IDL-SEBA candidate at time \(t\). The top row shows these instantaneous supports \(A_j(t)\), while the bottom row shows the advected images \(F_0^t(A_j(0))\), obtained by transporting the pixels belonging to the initial support at \(t=0\).

The comparison provides a diagnostic of quasi-materiality. The IDL supports remain close to the independently advected sets over the interval, but they are not constrained to coincide with them exactly. In this example, the extracted structures deform and contract with the compressible photospheric flow while retaining coherent organization over the analysis window.

\begin{figure}[t!]
    \centering
    \includegraphics[width=0.8\textwidth]{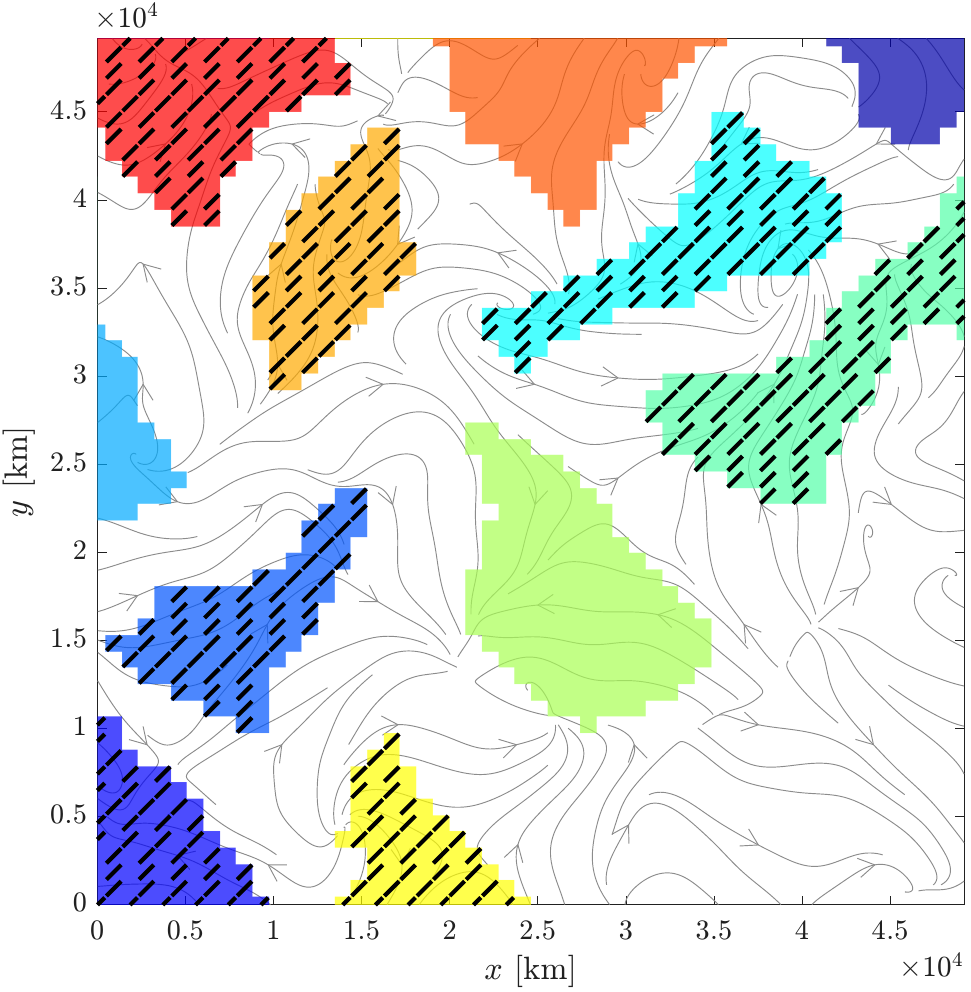}
    \caption{IDL-SEBA coherent-set candidates and LAVD-based classification at \(t=0\). Hatched regions indicate candidates selected by the LAVD criterion.}
    \label{fig:ftrcs}
\end{figure}

The IDL construction identifies finite-time coherent sets, but these sets are not necessarily rotational. To isolate finite-time rotationally coherent structures, we use LAVD as an independent post-classification diagnostic. The LAVD field is not used in the IDL computation or in the SEBA extraction. Instead, for each IDL-SEBA candidate we consider its initial support \(A=A_j(0)\subset M\) and evaluate the rotational enrichment
\begin{equation}
    E_{\rm LAVD}
    :=
    \frac{
    \operatorname{area}(A)^{-1}\int_A {\rm LAVD}_0^{6\,{\rm h}}(x)\,{\rm d}x
    }
    {
    \operatorname{area}(M)^{-1}\int_M {\rm LAVD}_0^{6\,{\rm h}}(x)\,{\rm d}x
    } .
\end{equation}
Candidates with \(E_{\rm LAVD}>1\) have mean accumulated intrinsic rotation above the domain average and are classified as rotationally enriched IDL coherent sets. Figure~\ref{fig:ftrcs} shows the IDL-SEBA candidates at \(t=0\), with black hatching indicating the candidates that satisfy this LAVD criterion. These structures should not be interpreted as structures obtained solely from the LAVD field. Instead, they are transport-coherent IDL candidates that also exhibit enhanced intrinsic rotation. The construction is related to the recently introduced Lagrangian rotating coherent structures (LRCS) for compressible flows \cite{Beron-26-Chaos}, which identify material regions that combine enhanced intrinsic rotation, measured by LAVD, with finite-time material contraction.  Here, the organizing material property is instead provided by finite-time transport coherence identified by the IDL, after which the rotational content is evaluated using LAVD.

To quantify these properties, Table~\ref{tab:ftrcs} reports the relative contributions
\begin{equation}
    r_{\rm space}
    :=
    \frac{\rho_{\rm space}}{\rho_a},
    \qquad
    r_{\rm material}
    :=
    \frac{a^2\rho_{\rm material}}{\rho_a},
\end{equation}
which measure the fractions of the IDL-SEBA Rayleigh quotient associated with spatial compactness and material persistence, respectively. The table also reports \(E_{\rm LAVD}\), allowing the material coherence and rotational enrichment diagnostics to be compared candidate by candidate.

\begin{table}[t!]
    \centering
    \begin{tabular}{cccc}
        \toprule
        Set & \(r_{\rm space}\) & \(r_{\rm material}\) & \(E_{\rm LAVD}\) \\
        \midrule
        \textcolor[rgb]{0.300,0.300,0.767}{\(\blacksquare\)}
            & 0.620 & 0.380 & 1.080 \\
        \textcolor[rgb]{0.300,0.300,1.000}{\(\blacksquare\)}
            & 0.747 & 0.253 & 0.944 \\
        \textcolor[rgb]{0.300,0.533,1.000}{\(\blacksquare\)}
            & 0.706 & 0.294 & 1.465 \\
        \textcolor[rgb]{0.300,0.767,1.000}{\(\blacksquare\)}
            & 0.620 & 0.380 & 0.954 \\
        \textcolor[rgb]{0.300,1.000,1.000}{\(\blacksquare\)}
            & 0.664 & 0.336 & 0.868 \\
        \textcolor[rgb]{0.533,1.000,0.767}{\(\blacksquare\)}
            & 0.516 & 0.484 & 0.871 \\
        \textcolor[rgb]{0.767,1.000,0.533}{\(\blacksquare\)}
            & 0.662 & 0.338 & 1.123 \\
        \textcolor[rgb]{1.000,1.000,0.300}{\(\blacksquare\)}
            & 0.429 & 0.571 & 0.832 \\
        \textcolor[rgb]{1.000,0.767,0.300}{\(\blacksquare\)}
            & 0.451 & 0.549 & 0.796 \\
        \textcolor[rgb]{1.000,0.533,0.300}{\(\blacksquare\)}
            & 0.269 & 0.731 & 1.067 \\
        \textcolor[rgb]{1.000,0.300,0.300}{\(\blacksquare\)}
            & 0.988 & 0.012 & 0.857 \\
        \bottomrule
    \end{tabular}
    \caption{Diagnostics of the IDL-SEBA coherent-set candidates shown in Fig.~\ref{fig:ftrcs}.}
    \label{tab:ftrcs}
\end{table}

As an independent check of the rotational interpretation, we also considered the deformation of material markers initialized inside each IDL-SEBA candidate. For each set, a circular marker with a material spoke was initialized at \(t=0\) and advected by the observed velocity field. The angular displacement of the spoke was then computed as
\begin{equation}
    \Delta\theta(t)
    :=
    \arg\left(
    x_{\rm tip}(t)-x_{\rm c}(t)
    \right)
    -
    \arg\left(
    x_{\rm tip}(0)-x_{\rm c}(0)
    \right),
\end{equation}
where \(x_{\rm c}(t)\) and \(x_{\rm tip}(t)\) denote the advected center and tip of the material spoke. This diagnostic is not used in the selection procedure, but provides a direct visualization of short-time material rotation within the IDL candidates. Unlike the LAVD enrichment, which measures accumulated intrinsic rotation averaged over the entire coherent set, the spoke angle follows a particular material marker and can therefore indicate localized rotation within a candidate.

\begin{figure}[t!]
    \centering

    \includegraphics[width=\textwidth]{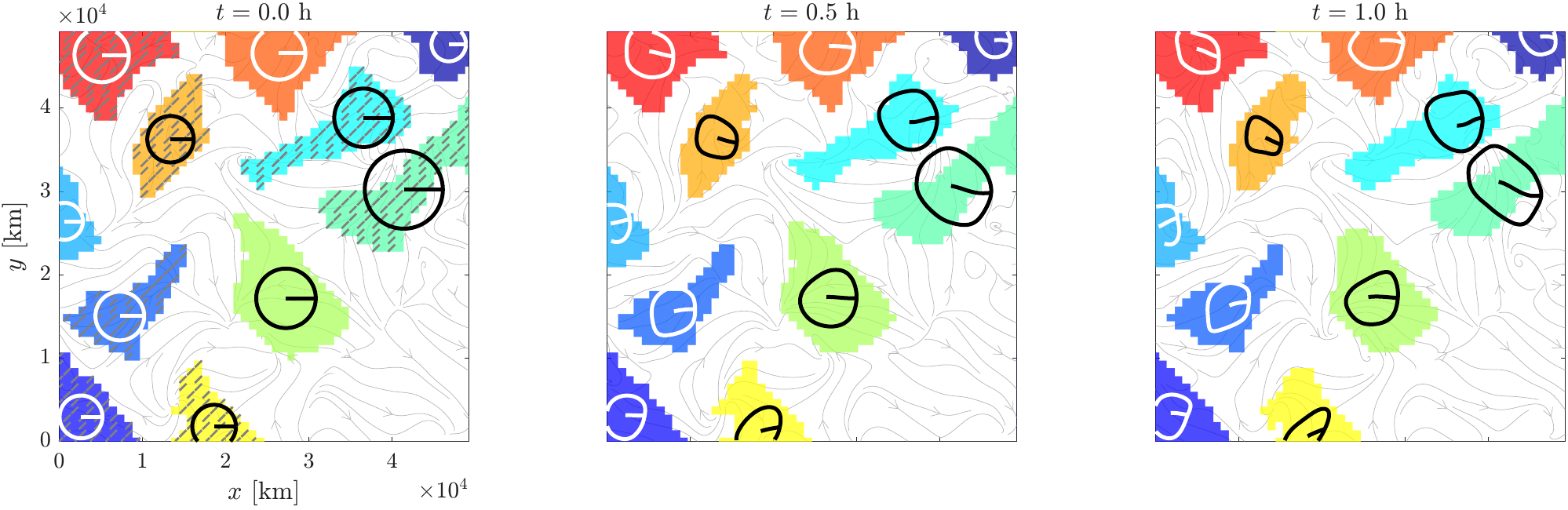}

    \vspace{0.2cm}

    \includegraphics[width=0.8\textwidth]{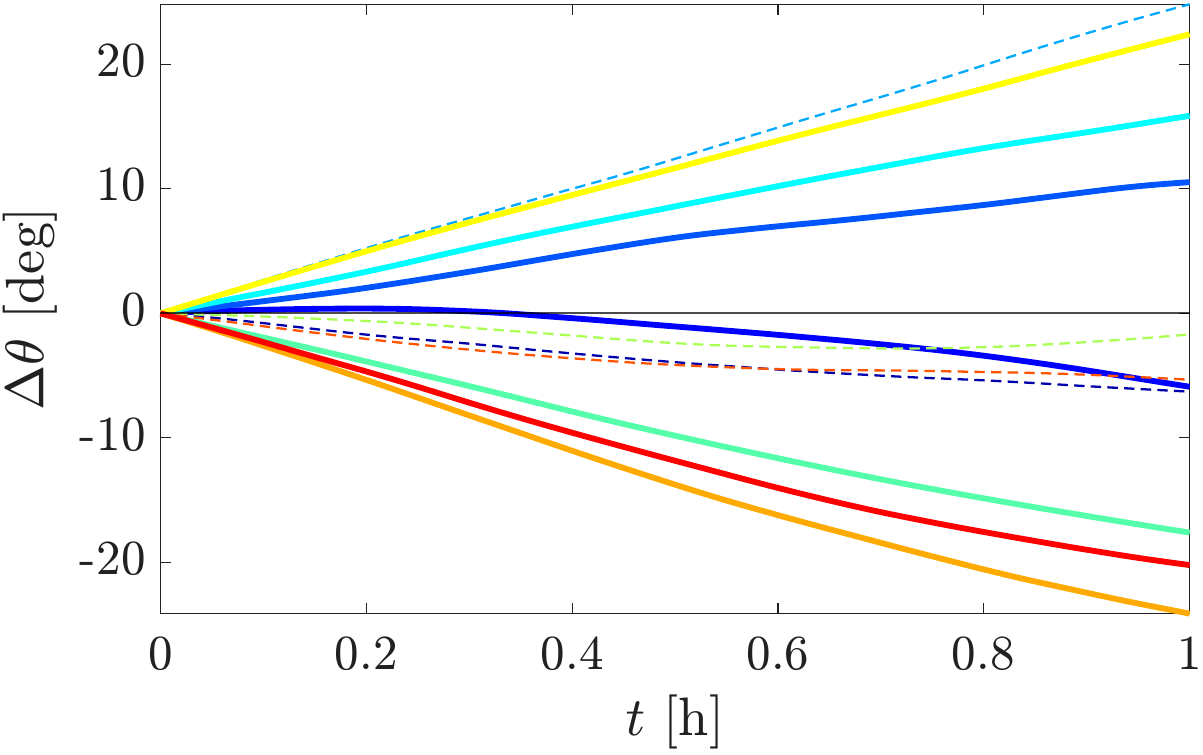}

    \caption{Material-marker rotation diagnostic for the IDL-SEBA candidates. Top: evolution of advected loop-and-spoke markers initialized inside the candidates. Bottom: angular displacement of the material spokes; solid curves correspond to LAVD-selected candidates and dashed curves to the remaining IDL-SEBA candidates.}
    \label{fig:rotation_marker}
\end{figure}

Figure~\ref{fig:rotation_marker} shows the evolution of the material markers over the first hour of the interval and the corresponding angular displacement. The LAVD-selected candidates generally display coherent material rotation, whereas several other IDL coherent sets primarily undergo translation and deformation. One non-selected candidate exhibits a relatively large spoke rotation despite having \(E_{\rm LAVD}<1\), illustrating that localized marker rotation and set-averaged LAVD enrichment measure different aspects of the flow. This distinction is the reason why the marker diagnostic is used only for visualization, while the FTRCS classification is based on the LAVD enrichment of the full IDL candidate.

Finally, we examine the dependence of the extracted structures on the finite-time analysis window. This issue is particularly relevant for compressible flows. In nearly incompressible applications, the finite-lifetime capability of IDL can identify coherent structures undergoing topological changes, such as the splitting and merging of the stratospheric polar vortex \cite{Atnip-etal-24}. In such settings, coherent material regions approximately preserve area, and changes in the IDL space-time support primarily reflect the reorganization of transport barriers. In contrast, the horizontal photospheric flow considered here is strongly compressible, so coherent organization can also arise from persistent convergence and contraction of material regions.

\begin{figure}[t!]
    \centering
    \begin{subfigure}{0.49\textwidth}
        \includegraphics[width=\textwidth]{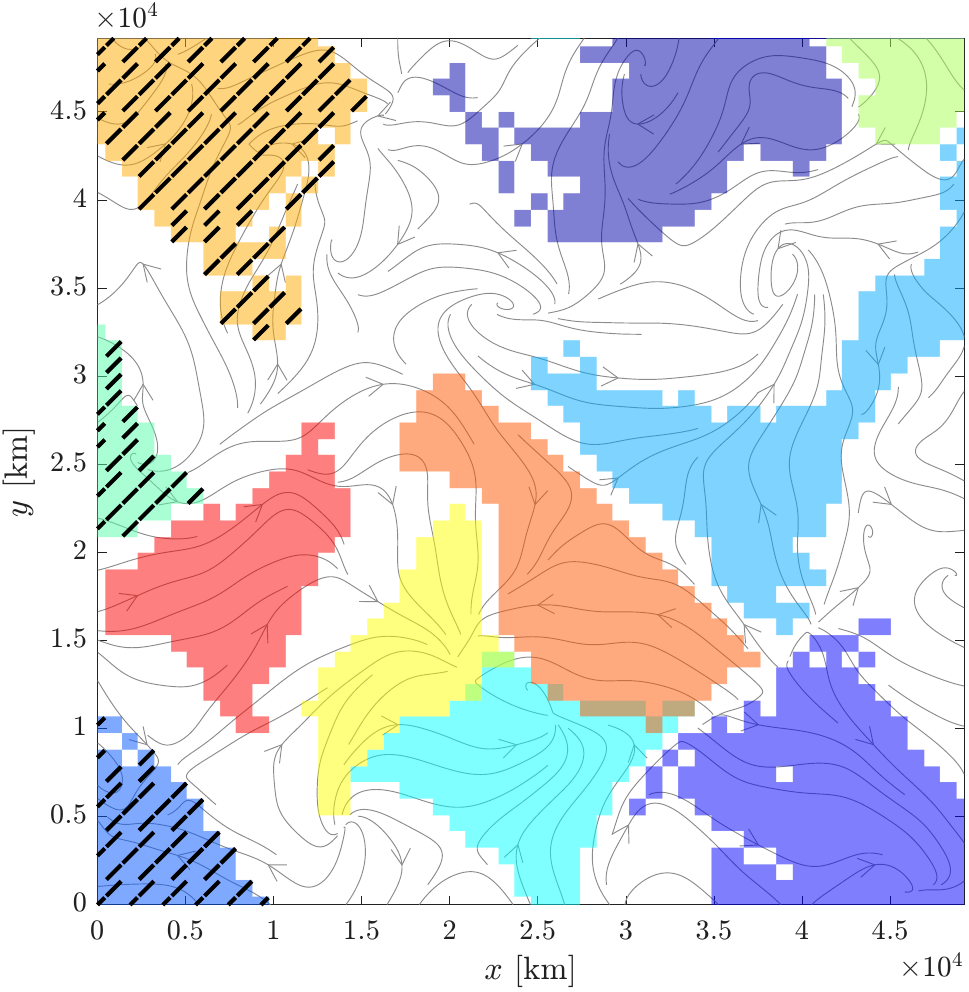}
        \caption{}
    \end{subfigure}
    \begin{subfigure}{0.49\textwidth}
        \includegraphics[width=\textwidth]{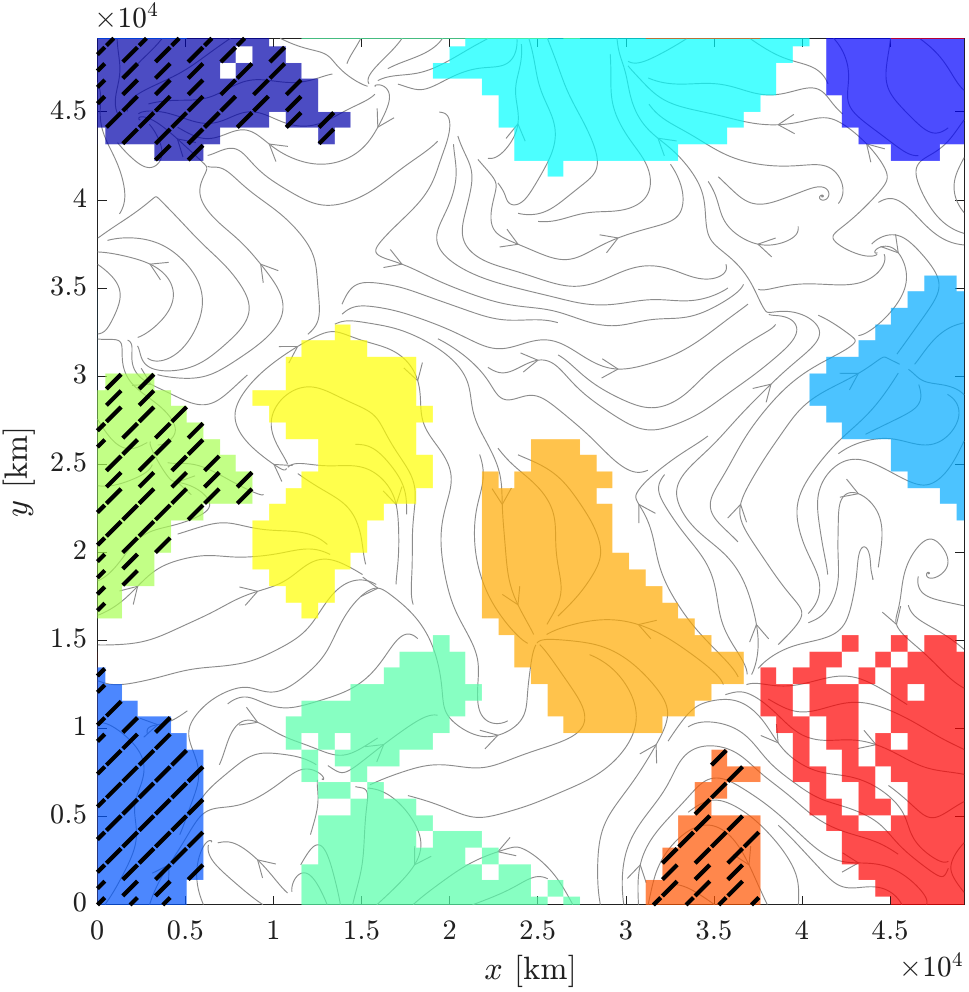}
        \caption{}
    \end{subfigure}
    \caption{Dependence of IDL-SEBA coherent-set candidates and LAVD-based rotational enrichment on the finite-time analysis window. Panel (a) uses the interval \(0\)--\(12\,{\rm h}\); panel (b) uses the shifted interval \(6\)--\(12\,{\rm h}\). Hatched regions indicate IDL candidates satisfying the LAVD enrichment criterion.}
    \label{fig:idl_windows}
\end{figure}

Figure~\ref{fig:idl_windows} compares IDL-SEBA candidates obtained over the intervals \(0\le t\le12\,{\rm h}\) and \(6\le t\le12\,{\rm h}\), using the same inflation parameter \(a=0.03\). The hatching indicates the subset of candidates satisfying the LAVD enrichment criterion and therefore classified as FTRCS. The longer calculation identifies coherent regions whose evolution reflects the organization accumulated over the full observation window, including the effect of persistent contraction by the convergent component of the supergranular flow. Restarting the analysis at \(t=6\,{\rm h}\) produces a different collection of coherent sets associated with the transport organization present during the later part of the evolution. Some of these later-time coherent sets also satisfy the rotational enrichment criterion, reflecting the renewal of finite-time rotational organization. Thus, in compressible flows, extending the analysis interval and shifting the finite-time window provide complementary information: the former emphasizes coherence over a prescribed time horizon, whereas the latter reveals the emergence and disappearance of coherent structures as the flow evolves.

\section{Concluding remarks}\label{lab:conclusions}

This work introduced finite-time rotationally coherent sets (FTRCS) for the analysis of photospheric supergranulation. The construction separates two aspects of coherent motion: finite-time quasi-material coherence, identified using the inflated dynamic Laplacian (IDL), and intrinsic material rotation, quantified independently using the Lagrangian-averaged vorticity deviation (LAVD).

Application to photospheric velocity data showed that instantaneous vortical features do not necessarily determine finite-time rotational coherence. The IDL identifies coherent transport regions with finite lifetimes, while the subsequent LAVD evaluation determines which of these regions exhibit enhanced intrinsic rotation. Rotational coherence is therefore a more restrictive property than quasi-material transport coherence.

The analysis also illustrates the role of compressibility in the interpretation of finite-time coherent structures. In photospheric supergranulation, coherent regions may arise from contraction associated with convergent transport rather than from rotating material regions with approximately preserved area. Comparing different analysis windows shows how the identified structures depend on the selected finite-time interval and provides information about their formation and disappearance.

FTRCS are therefore not necessarily material vortices in the classical sense, but finite-time quasi-material coherent regions with enhanced intrinsic rotation. The same approach can be applied to other time-dependent flows in which coherent structures form, deform, contract, or disappear.

\section*{Acknowledgments}

F.J.B.-V.\ thanks Suzana S.A.\ Silva and Erico L.\ Rempel for insightful discussions on photospheric supergranulation flows, and S.S.A.\ Silva for facilitating access to the processed Hinode velocity data set. F.J.B.-V.\ thanks Milan Go\v{s}i\'{c} for permission to use this data product. F.J.B.-V.\ also thanks the International Space Science Institute (ISSI) in Bern, Switzerland, for the hospitality provided to the members of the team ``Opening new avenues in identifying coherent structures and transport barriers in the magnetised solar plasma.''

\section*{Author Declarations}

\subsection*{Funding}

F.J.B.-V.\ received no external funding for this work.

\subsection*{Conflict of Interest}

F.J.B.-V.\ has no conflicts of interest to disclose.

\subsection*{Author Contributions}

F.J.B.-V.\ conceived the study, developed the methodology, carried out the numerical analysis, generated the figures, and wrote the manuscript. ChatGPT was used as an assistance tool for code development, debugging, and editorial refinement.

\appendix

\numberwithin{equation}{section}
\renewcommand{\theequation}{\Alph{section}.\arabic{equation}}

\section{Dynamic Laplacian and material coherent sets}
\label{app:dl}

Let \(M\) denote the two-dimensional flow domain at the initial time \(t_0\), and let \(F_{t_0}^{t}:M\to M_t\), \(x\mapsto x_t\), be the flow map generated by the velocity field \(\mathbf u(x,t)\). A material set \(A\subset M\) evolves under the dynamics into the set \(F_{t_0}^{t}A\) at time \(t\).

The deformation induced by the flow can be described geometrically by pulling back the Euclidean metric from the evolved domain \(M_t\) to the reference configuration \(M\). In Cartesian coordinates, this pullback metric is represented by the right Cauchy--Green tensor \cite{Haller-Beron-12},
\begin{equation}
   C_{t_0}^{t}(x)
   :=
   \nabla F_{t_0}^{t}(x)^\top
   \nabla F_{t_0}^{t}(x).
\end{equation}
The tensor \(C_{t_0}^{t}\) measures the finite-time deformation of infinitesimal material vectors and therefore determines how material boundaries stretch under the flow.

The central idea behind finite-time coherent sets is that a coherent material region should resist filamentation: its boundary should remain relatively small compared with its size throughout the observation interval. This places coherent-set detection in the class of isoperimetric problems.

The classical isoperimetric problem seeks, among all planar sets \(A\) of prescribed area, the one with minimal boundary length \cite{Osserman-78}. Introducing a Lagrange multiplier for the area constraint gives the variational problem
\begin{equation}
   \delta
   \big(
   |\partial A|
   +
   \lambda |A|
   \big)
   =
   0 ,
\end{equation}
where, to simplify notation, \(|\cdot|\) denotes the corresponding geometric measure: length when applied to curves and area when applied to planar sets. If the boundary is perturbed by a normal displacement \(\eta\), the first variations are
\begin{equation}
   \delta |\partial A|
   =
   \oint_{\partial A}
   \kappa\eta\,ds,
   \qquad
   \delta |A|
   =
   \oint_{\partial A}
   \eta\,ds ,
\end{equation}
where \(\kappa\) is the signed curvature of the boundary \(\partial A\). Hence stationarity requires
\begin{equation}
   \kappa+\lambda=0 ,
\end{equation}
showing that an extremizing boundary must have constant curvature. The resulting closed planar curve is a circle, which gives the solution of the classical isoperimetric problem.

Finite-time coherent sets generalize this geometric idea by replacing the length of the initial boundary with the average length of its material images under the flow. Froyland \cite{Froyland-15} introduced the dynamic Cheeger ratio
\begin{equation}
   h^D(A)
   :=
   \frac{
      \overline{|\partial F_{t_0}^{t}A|}^{t_0,t_1}
   }
   {
      \min\left\{
      |A|,
      |M\!\setminus\!A|
      \right\}
   } .
\end{equation}
Using the pullback metric induced by the flow map, the numerator can be expressed entirely on the reference domain as
\begin{equation}
   h^D(A)
   =
   \frac{
      \frac{1}{T}
      \int_{t_0}^{t_1}
      \int_{\partial A}
      \sqrt{dx^\top C_{t_0}^{t}(x)dx}
      \,dt
   }
   {
      \min\left\{
      \int_A d^2x,
      \int_{M\setminus A}d^2x
      \right\}
   },
   \qquad
   T=t_1-t_0 .
\end{equation}

Thus the problem is still an isoperimetric problem over sets, but now the relevant boundary length is measured in a time-averaged geometry induced by the dynamics. A formal first variation again leads to an Euler--Lagrange condition involving a generalized curvature; however, this curvature is now associated with the averaged pullback metric rather than the Euclidean metric. Unlike the classical problem, the resulting equation is flow-dependent, nonlocal in time, and does not yield explicit solutions except in special cases.

Minimizing \(h^D(A)\) identifies material sets whose average evolved boundary length remains small relative to their area. This criterion directly targets resistance to filamentation rather than small trajectory separation. Thus coherent sets identified through this variational principle need not consist of trajectories that remain close together, but rather of material regions whose collective geometry is preserved.

Because the dynamic Cheeger ratio is expressed in terms of intrinsic geometric quantities of material sets and their advected images, it is unchanged by time-dependent translations and rotations of the observer frame. It therefore defines an objective measure of finite-time material coherence.

For compressible flows, the same geometric principle applies, but the measure used to quantify the size of a material set must be specified. If an advected density is relevant, then the denominator of the Cheeger ratio should be modified to use the corresponding material mass. In this work, we use the reference area measure because the objective is to identify coherent transport regions in the observed horizontal velocity field. Compressibility is therefore retained through the deformation of material boundaries encoded by the flow map.

The minimization of \(h^D(A)\) is difficult because the unknown is the set \(A\) itself. After discretization, this leads to a combinatorial partitioning problem. The dynamic Laplacian provides a spectral relaxation by replacing the search over characteristic functions of sets with an eigenvalue problem. With the convention that the Laplacian is negative semidefinite,
\begin{equation}
   \Delta^D f_i
   =
   \lambda_i f_i,
   \qquad
   0=\lambda_1>\lambda_2\ge\lambda_3\ge\cdots .
\end{equation}

The dynamic Laplacian is obtained by averaging Laplace operators transported by the flow,
\begin{equation}
   \Delta^D f
   :=
   \frac1T
   \int_{t_0}^{t_1}
   \left(F_{t_0}^{t}\right)^*
   \Delta
   \left(F_{t_0}^{t}\right)_*
   f
   \,dt ,
\end{equation}
where \((F_{t_0}^{t})_*\) and \((F_{t_0}^{t})^*\) denote push-forward and pull-back operations. In coordinates this becomes
\begin{equation}
   \Delta^D f
   =
   \frac1T
   \int_{t_0}^{t_1}
   \frac{1}{\det(C_{t_0}^{t})^{1/2}}
   \nabla\cdot
   \left(
   \det(C_{t_0}^{t})^{1/2}
   C_{t_0}^{t\, -1}
   \nabla f
   \right)
   dt .
\end{equation}
The dynamic Laplacian can therefore be interpreted as an elliptic operator on the reference domain whose coefficients encode the average deformation experienced by material gradients.

The connection between the dynamic Laplacian and the dynamic isoperimetric problem is provided by dynamic analogues of classical Cheeger inequalities \cite{Cheeger-70, Froyland-15}. In the incompressible case, the second eigenvalue satisfies bounds of the form
\begin{equation}
   \frac14
   \left(
   \inf_A h^D(A)
   \right)^2
   \le
   -\lambda_2
   \le
   c\,\inf_A h^D(A),
\end{equation}
where \(c\) depends on the geometric normalization. Thus an eigenvalue close to zero implies the existence of a material partition with a small average boundary-to-area ratio, and conversely a small dynamic Cheeger constant implies a small value of \(-\lambda_2\).

The same interpretation follows from the Rayleigh quotient. For an eigenfunction \(f_i\),
\begin{equation}
   -\lambda_i
   =
   \frac{
      \frac1T
      \int_{t_0}^{t_1}
      \int_M
      \nabla f_i^\top
      C_{t_0}^{t\, -1}
      \nabla f_i
      \det(C_{t_0}^{t})^{1/2}
      \,d^2x\,dt
   }
   {
      \int_M f_i^2\,d^2x
   } .
\end{equation}
An eigenvalue close to zero therefore indicates that \(f_i\) has small average Dirichlet energy in the family of geometries induced by the flow. In practical terms, \(f_i\) varies slowly within coherent regions and changes mainly across their boundaries. Level sets or localized combinations of eigenfunctions can then be used to approximate coherent material sets.

For numerical computation, the weak formulation is discretized using the finite element method (FEM) \cite{Froyland-Junge-18}. The \(i\)-th eigenfunction is approximated as
\begin{equation}
   f_i(x)
   \approx
   \sum_{j=1}^{N}
   (v_i)_j\phi_j(x),
\end{equation}
where \(\{\phi_j\}\) is a finite-element basis. The coefficient vectors satisfy the generalized eigenvalue problem
\begin{equation}
   Kv_i
   =
   \mu_iBv_i,
   \qquad
   \mu_i=-\lambda_i\ge0 .
\end{equation}
Here \(B\) is the mass matrix,
\begin{equation}
   B_{ij}
   =
   \int_M
   \phi_i\phi_j\,d^2x ,
\end{equation}
and \(K\) is the dynamic stiffness matrix,
\begin{equation}
   K_{ij}
   =
   \frac1T
   \int_{t_0}^{t_1}
   \int_M
   \nabla\phi_i^\top
   C_{t_0}^{t\, -1}
   \nabla\phi_j
   \det(C_{t_0}^{t})^{1/2}
   \,d^2x\,dt .
\end{equation}

The dynamic Laplacian therefore provides a computational relaxation of the finite-time material isoperimetric problem. The inflated dynamic Laplacian used in this work extends this construction by replacing the search for a single material set over the full interval with a search for coherent space-time structures whose lifetimes are determined as part of the optimization.

\section{Inflated dynamic Laplacian and quasi-material coherent sets}
\label{app:idl}

The dynamic Laplacian formulation described in Appendix~\ref{app:dl} identifies material coherent sets: a single initial set \(A\subset M\) is selected such that its images \(F_{t_0}^{t}A\) remain coherent throughout the prescribed interval \([t_0,t_1]\). This requirement is appropriate for persistent material structures, but it can be restrictive in flows where coherent regions have finite lifetimes. In such cases, coherent structures may appear, disappear, merge, split, or undergo substantial changes during the observation window.

The inflated dynamic Laplacian (IDL) of \cite{Froyland-Koltai-23} relaxes this restriction by changing the object being optimized. Instead of searching for a material set \(A\), one searches for a coherent set in space-time,
\begin{equation}
   \mathcal A
   \subset
   \bigcup_{t\in[t_0,t_1]}
   M_t\times\{t\}.
\end{equation}
The time slices \(\mathcal A(t)\subset M_t\) describe the instantaneous spatial support of the structure. Classical material coherent sets correspond to the special case
\begin{equation}
   \mathcal A(t)
   =
   F_{t_0}^{t}A ,
\end{equation}
for some fixed initial set \(A\subset M\). More generally, the IDL identifies \emph{quasi-material} coherent sets whose evolution remains close to material evolution while allowing their spatial support to vary within the observation window.

The construction is based on defining a geometry on the space-time domain obtained after pulling the dynamics back to the reference configuration. Spatial directions inherit the same pullback geometry used by the dynamic Laplacian, while the time direction is assigned a length scale through an inflation parameter \(a\). Formally, the inflated metric can be viewed as augmenting spatial distances by
\begin{equation}
   |dx_t|^2+a^{-2}dt^2 .
\end{equation}
The parameter \(a\) converts time into a spatial length scale and controls the relative cost of temporal boundaries.

Small values of \(a\) make temporal changes inexpensive, allowing structures to have shorter lifetimes or stronger deviations from material evolution. Increasing \(a\) penalizes variations along trajectories and favors space-time structures that persist under the flow. In the large-\(a\) limit, the optimization approaches the material coherent-set problem associated with the dynamic Laplacian.

The spectral relaxation of the inflated isoperimetric problem leads to an eigenvalue problem on the augmented space-time domain. In the negative-semidefinite convention,
\begin{equation}
   \Delta_a^D f_i
   =
   \lambda_i(a)f_i .
\end{equation}
The inflated operator may be written formally as
\begin{equation}
   \Delta_a^D
   :=
   \Delta^D
   +
   a^2\partial_{tt},
\end{equation}
where the time derivative represents variation along the material direction in the inflated geometry.

Equivalently, the variational interpretation is expressed through the Rayleigh quotient
\begin{equation}
   -\lambda_i(a)
   =
   \frac{
      \int_{t_0}^{t_1}
      \int_{M_t}
      |\nabla f_i|^2
      +
      a^2
      \left(
      \frac{Df_i}{Dt}
      \right)^2
      d^2x_t\,dt
   }
   {
      \int_{t_0}^{t_1}
      \int_{M_t}
      f_i^2\,d^2x_t\,dt
   },
\end{equation}
where
\begin{equation}
   \frac{D}{Dt}
   =
   \partial_t+\mathbf u\cdot\nabla
\end{equation}
is the material derivative. The first contribution measures spatial variation within instantaneous time slices, whereas the second measures variation along trajectories. Therefore, eigenfunctions associated with eigenvalues close to zero identify functions that remain nearly constant inside coherent space-time tubes.

The numerical implementation follows from a finite-element discretization of the inflated weak formulation. Expanding the eigenfunctions in a space-time finite-element basis gives the generalized eigenvalue problem
\begin{equation}
   K(a)v_i
   =
   \mu_i(a)Bv_i,
   \qquad
   \mu_i(a)=-\lambda_i(a)\ge0 .
\end{equation}
The stiffness matrix separates naturally into spatial and material contributions,
\begin{equation}
   K(a)
   =
   K_{\rm space}
   +
   a^2K_{\rm material}.
\end{equation}
The first term penalizes spatial gradients within individual time slices, while the second penalizes variation along material trajectories.

For any discrete function represented by coefficient vector \(w\), the corresponding Rayleigh quotient is
\begin{equation}
   \rho_a(w)
   :=
   \frac{w^\top K(a)w}{w^\top Gw}
   =
   \frac{w^\top K_{\rm space}w}{w^\top Gw}
   +
   a^2
   \frac{w^\top K_{\rm material}w}{w^\top Gw}.
\end{equation}
Here \(G\) is the mass (Gram) matrix defining the discrete \(L^2\) inner product. Thus the discrete IDL optimization balances two competing effects: minimizing instantaneous spatial boundaries and maintaining coherence along the flow.

The low-dimensional eigenspace associated with the smallest nonzero eigenvalues contains the dominant coherent space-time partitions. Since individual eigenfunctions need not correspond to isolated structures, we apply sparse eigenbasis approximation (SEBA) to obtain localized combinations of these modes. The resulting functions provide candidate coherent sets whose supports are extracted by thresholding.

For each candidate set, the two contributions to \(\rho_a\) provide diagnostic information. The spatial contribution measures compactness of the space-time support, whereas the material contribution measures departure from material evolution. The parameter \(a\) is therefore interpreted as a scale parameter rather than as an optimization variable with a universal best value. In this work, we examine the dependence of the eigenvalues and Rayleigh contributions on \(a\) to identify regimes in which the selected structures balance spatial coherence and material persistence.

The IDL therefore generalizes the dynamic Laplacian from persistent material sets to finite-lifetime quasi-material structures. These structures form the transport-coherent candidates that are subsequently tested for enhanced intrinsic rotation using LAVD.

\bibliographystyle{plain}
\bibliography{fot}

\subsection*{Data and Software Availability}

The Hinode observations used here were acquired in the framework of Hinode Operation Plan 151, ``Flux replacement in the solar network and internetwork.'' The underlying Hinode/SOT observations are publicly available through the Hinode data archive (\href{https://darts.isas.jaxa.jp/en/missions/hinode}{https://darts.isas.jaxa.jp/en/missions/hinode}). The processed photospheric velocity data set used in this study was provided by S.S.A.\ Silva and used with permission from M. Go\v{s}i\'{c}. Requests for access to this processed data product should be directed to the original providers. The MATLAB codes used to generate the figures and results presented in this paper are available from F.J.B.-V.\ upon reasonable request.

\end{document}